\documentclass[a4paper,final,conference]{IEEEtran}
\usepackage{mathpple}
\usepackage{times}

\usepackage[top=0.58in, bottom=0.58in, left=0.6in, right=0.6in]{geometry}

\usepackage{amsmath}  
\usepackage{amssymb}  
\usepackage{mathrsfs} 
\usepackage{kantlipsum,cuted}
\usepackage{multirow}
\usepackage{tabu}

\usepackage{lipsum}

\usepackage{theorem}  
\usepackage{cite}     
\usepackage{comment}  

\usepackage{upref}
\usepackage{amsfonts}

\usepackage{verbatim}

\usepackage[dvipsnames,usenames]{color}
\usepackage{enumerate}
\usepackage{multicol}
\usepackage{blkarray, bigstrut}
\usepackage{graphicx}
\usepackage{subfigure}

\usepackage{latexsym}

\usepackage[draft,ulem=normalem]{changes}
\usepackage[all]{xy}

\usepackage{algorithmicx}
\usepackage{algorithm}
\usepackage[noend]{algpseudocode}
\algrenewcommand\alglinenumber[1]{\scriptsize #1:}
\algrenewcommand\algorithmicindent{1em}%

\usepackage{wrapfig}

\makeatletter
\def\BState{\State\hskip-\ALG@thistlm}

\makeatother

\newcommand{\be}[1]{\begin{equation}\label{#1}}
\newcommand{\ee}{\end{equation}}

\newcommand{\bc}{\begin{center}}
\newcommand{\ec}{\end{center}}

\newcommand{\cC}{{\cal C}}

\newcommand{\cG}{{\cal G}}
\newcommand{\cH}{{\cal H}}

\newcommand{\cO}{{\cal O}}

\newcommand{\cS}{{\cal S}}

\newcommand{\bfc}{{\boldsymbol c}}

\newcommand{\bfg}{{\boldsymbol g}}

\renewcommand{\leq}{\leqslant}

\renewcommand{\geq}{\geqslant}

\newcommand{\F}{\mathbb{F}}
\newcommand{\Fq}{\smash{\mathbb{F}_{\!q}}}

\newcommand{\Cref}[1]{Co\-rol\-la\-ry\,\ref{#1}}

\theoremstyle{plain} \theorembodyfont{\normalfont\slshape}

\newtheorem{thm}{Theorem$\!$}
\newenvironment{theorem}{\begin{thm}\hspace*{-1ex}{\bf.}}{\end{thm}}

\newtheorem{prop}[thm]{Proposition$\!$}

\newtheorem{lem}[thm]{Lemma$\!$}
\newenvironment{lemma}{\begin{lem}\hspace*{-1ex}{\bf.}}{\end{lem}}

\newtheorem{cor}[thm]{Corollary$\!$}

\newtheorem{prob}[thm]{Problem$\!$}

\newtheorem{defi}[thm]{Definition$\!$}
\newenvironment{definition}{\begin{defi}\hspace*{-1ex}{\bf.}}{\end{defi}}

\newtheorem{claim}{Claim}

\theorembodyfont{\normalfont}

\newtheorem{exam}{Example$\!$}
\newenvironment{example}{\begin{exam}\hspace*{-1ex}{\bf .}}{\end{exam}}

\newtheorem{remrk}{Remark$\!$}

\newtheorem{Construction}{Construction}

\definecolor{Codecolor}{named}{White}  

\newcommand{\Copen}{\mbox{\{\kern-5.50pt\{}}
\newcommand{\Cclose}{\mbox{\}\kern-5.50pt\}}}
\newcommand{\Cslash}{\mbox{$\backslash\kern-6.02pt\backslash$}}

\newcommand{\gc}{\cG\textmd{-}}

\begin{document}

\title{\textbf{Codes for Graph Erasures}\vspace{-1ex}}
\author{\IEEEauthorblockN{\textbf{Lev Yohananov}}
	\IEEEauthorblockA{Dept. of Computer Science\\
		Technion-Israel Institute of Technology\\
		Haifa 32000, Israel \\
		Email: levyohananov@campus.technion.ac.il\vspace{-4ex}}
	\and
	\IEEEauthorblockN{\textbf{Eitan Yaakobi}}
	\IEEEauthorblockA{Dept. of Computer Science\\
		Technion-Israel Institute of Technology\\
		Haifa 32000, Israel \\
		Email: yaakobi@cs.technion.ac.il\vspace{-4ex}}}
\maketitle
\begin{abstract}
Motivated by systems where the information is represented by a graph, such as neural networks, associative memories, and distributed systems, we present in this work a new class of codes, called \emph{codes over graphs}. Under this paradigm, the information is stored on the edges of an undirected graph, and a code over graphs is a set of graphs. A \emph{node failure} is the event where all edges in the neighborhood of the failed node have been erased. We say that a code over graphs can tolerate $\rho$ node failures if it can correct the erased edges of any $\rho$ failed nodes in the graph. While the construction of such codes can be easily accomplished by MDS codes, their field size has to be at least $\cO(n^2)$, when $n$ is the number of nodes in the graph. In this work we present several constructions of codes over graphs with smaller field size. In particular, we present optimal codes over graphs correcting two node failures over the binary field, when the number of nodes in the graph is a prime number. We also present a construction of codes over graphs correcting $\rho$ node failures for all $\rho$ over a field of size at least $(n+1)/2-1$, and show how to improve this construction for optimal codes when $\rho=2,3$.
\end{abstract}\vspace{-1ex}

\section{Introduction}\vspace{-0ex}
\renewcommand{\baselinestretch}{0.97}\normalsize\noindent

The traditional setup to represent information is by a vector over some fixed alphabet. Although this commonly used model is the most practical one, especially for storage and communication applications, it does not necessarily fit all information systems. In this work we study a different approach where the information is represented by a \emph{graph}. This model is  motivated by several information systems. For example, in \emph{neural networks}, the neural units are connected via \emph{links} which store and transmit information between the neural units~\cite{Hopfield:1988:NNP:65669.104422}. Similarly, in associative memories, the information is stored by associations between different data items~\cite{6283016}. These two examples mimic the brain functionality which stores and processes information by associations between the information units. Furthermore, representing information in a graph can model a distributed storage systems~\cite{NetworkCoding} while every two nodes can share a link with the information that is stored between the nodes.

In this paper we present a new class of codes which we call \emph{codes over graphs}. Under this setup we assume that there is an undirected complete graph with $n$ \emph{nodes} (vertices) such that the information is stored on the edges connecting between every two nodes in the graph, including self loops. The information on each edge is a symbol over some fixed alphabet so every graph can be represented by the symbols stored in each of the $\binom{n+1}{2}$ edges, and a code over graphs is simply a set of graphs. A \emph{node failure} is the event where all edges in the node's neighborhood have been erased, and the goal of this work is to construct codes over graphs that can efficiently correct node failures. Namely, we say that a code over graphs can correct $\rho$ node failures if it is possible to correct the erased edges in the neighborhoods of any failed $\rho$ nodes. We study node failures since they correspond to the events of failing neural units in a neural network, data loss in an associative memory, and unavailable and failed nodes in distributed storage systems. 

Since every graph can be represented by its adjacency matrix, a natural approach to construct codes over graphs is by their adjacency matrices. Thus, this class of codes is quite similar to the class of \emph{array codes}, such as maximum-rank array codes~\cite{DBLP:journals/tit/Roth91}, B codes~\cite{DBLP:journals/tit/XuBBW99}, EVENODD codes~\cite{DBLP:journals/tc/BlaumBBM95}, RDP code~\cite{RDP}, X-codes~\cite{DBLP:journals/tit/XuBBW99}, and regenerating codes~\cite{ButterflyISIT,Raviv16,DBLP:journals/tit/TamoWB13}. However, there are two main differences between classical array codes and codes over graphs. First, since the graphs are undirected, the matrices are symmetric and square. Second, a failure of the $i$th node in the graph corresponds to the failure of the $i$th row and $i$th column in the adjacency matrix. Most existing constructions of array codes  are not designed for symmetric or even square matrices. Furthermore, these constructions do not support the row-column failure model. The closest model to this setup is the one studied by Roth for crisscross errors~\cite{DBLP:journals/tit/Roth91}, in which a fixed number of rows and columns have failed. While it is possible to use some of the results from~\cite{DBLP:journals/tit/Roth91}, we will show in the paper that they do not provide codes over graphs with good parameters.

Assume a code over graphs with $n$ nodes such that every edge stores a symbol. If $\rho$ nodes have failed then the number of edges that were erased is 
\begin{equation}\label{eq:lb}
\binom{n+1}{2}-\binom{n-\rho+1}{2} = \rho n -\binom{\rho}{2}.
\end{equation}
Therefore, the number of \emph{redundancy edges} for every code which tolerates $\rho$ node failures is at least $\rho n -\binom{\rho}{2}$. A code over graphs which meets this lower bound on the number of redundancy edges will be called an \emph{optimal code over graphs}. Note that it is possible to construct optimal codes over graphs by using an $[\binom{n+1}{2}, \binom{n-\rho+1}{2},\rho n -\binom{\rho}{2}+1]$ MDS code. However, this will impose using a field of size at least $\binom{n+1}{2}-1=\Theta(n^2)$. Thus, the problem under this setup is the construction of such codes over smaller fields with optimal or close to optimal redundancy. 

The rest of this paper is organized as follows. In Section~\ref{sec:defs}, we formally define the graph model studied in this paper and some preliminary results. In Section~\ref{sec:double}, we present our main result in the paper of optimal binary codes over graphs correcting two node failures, when the number of nodes is prime. In Section~\ref{sec:multiple}, we extend our results for codes over graphs correcting arbitrary number of node failures over a field of size at least $(n+1)/2-1$. While this construction is almost optimal with respect to the bound in~(\ref{eq:lb}), we show how to improve it for optimal codes correcting two and three node failures. Lastly, in Section~\ref{sec:n-2}, we study the existence of codes over graphs correcting $n-2$ node failures. Due to the lack of space some proofs of the results in the paper are omitted.

\section{Definitions and Preliminaries}\label{sec:defs}
For a positive integer $n$, the set $\{0,1,\ldots,n-1\}$ will be denoted by $[n]$. For a prime power $q$, ${\Fq}$ is the finite field of size $q$. A linear code of length $n$, dimension $k$, and minimum distance $d$ over $\Fq$ will be denoted by $[n,k,d]_q$.

We will denote an undirected graph by $G=(V,E)$, where $V=\{v_0,v_1,\ldots,v_{n-1}\}$ is a set of $n$ nodes (vertices) and $E\subseteq V\times V$ is its edge set. By a slight abuse of notation, every (undirected) edge in the graph will be denoted by $(v_i,v_j)$ where the order in this pair does not matter, that is, the pair $(v_i,v_j)$ is identical to the pair $(v_j,v_i)$. When possible we will denote the edges by $(v_i,v_j)$, where $i\geq j$. We assume that the graph is simple in the sense that there are no parallel edges, however every node can have a self loop.

A graph $G$ over an alphabet $\Sigma$ is defined by a \emph{labeling function} $L:E\rightarrow \Sigma$ and will be denoted by $G=(V,E,L)$. In this work we will extend the definition of the labeling function over $V\times V$, i.e. $L:V\times V\rightarrow \Sigma$, and have $0\in \Sigma$ denote the case in which an edge does not exist. Hence, we can fully characterize the graph $G$ by its vertex set $V$ and labeling function $L$, and will denote it by $G(V,L)$, where we simply consider it as a complete graph with self loops. Under this setup, the \emph{adjacency matrix} of the graph $G$ is an $n\times n$ matrix $A_G=[a_{i,j}]^{n-1,n-1}_{i=0,j=0}$, where $a_{i,j}= L(v_i,v_j)$ for all $i,j\in[n]$. We will also use the \emph{lower-triangle-adjacency matrix} of $G$ to be the $n\times n$ matrix $A_G'=[a'_{i,j}]^{n-1,n-1}_{i=0,j=0}$ such that $a'_{i,j}=a_{i,j}$ if $i\geq j$ and otherwise $a'_{i,j}=0$. For $i\in[n]$, the \textit{neighborhood} of the $i$th node, denoted by $N_i$, is the set of edges connected to this node. Since we assumed the graph is complete, the neighborhood is simply the set $N_i=\{(v_i,v_j) | j\in[n] \}$. 

Let $\Sigma$ be a ring and $G_1$ and $G_2$ be two graphs over $\Sigma$ with the same nodes set $V$. The operator $"+"$ between $G_1$ and $G_2$ over $\Sigma$, is defined by $G_1 + G_2 = G_3$, where $G_3$ is the unique graph satisfying $A_{G_1} + A_{G_2} = A_{G_3}$.
Similarly, the operator $"\cdot"$ between $G_1$ and an element $\alpha \in \Sigma$, is denoted by $\alpha\cdot G_1 = G_3$, where $G_3$ is the unique graph satisfying $\alpha \cdot A_{G_1} = A_{G_3}$.\vspace{-1ex}
\begin{definition}
A \textit{\textbf{code over graphs}} of size $M$, length $n$ over $\Sigma$ is a set of undirected graphs $\mathcal{C}_{\cG}=\{G_{i} = (V_{n},L_{i})  |  i\in[M]\}$ over $\Sigma$ where $V_n =\{v_0,\ldots,v_{n-1}\}$. We denote such a code by $\gc(n,M)_{\Sigma}$ and in case $\Sigma =\{0,1\}$, it will simply be denoted by $\gc(n,M)$. The \textit{\textbf{dimension}} of a code over graphs $\mathcal{C}_{\cG}$ is $k_\cG= \log_{|\Sigma|}M$, the \textit{rate} is $R_\cG = k_\cG/{{n+1 \choose 2}}$, and the \textit{\textbf{redundancy}} is defined to be $r_\cG = {n+1 \choose 2} - k_\cG$.

A code over graphs $\mathcal{C}_{\cG}$ over a ring $\Sigma$ will be called \textit{\textbf{linear}} if for every $G_1,G_2 \in \mathcal{C}_{\cG}$ and $\alpha,\beta \in \Sigma$ it holds  that $\alpha G_1 + \beta G_2 \in \mathcal{C}_{\cG}$. We denote this family of codes over graphs by $\gc[n,k_\cG]_\Sigma$. 

A linear code over graphs whose first $k$ nodes contain the $\binom{k+1}{2}$ unmodified information symbols on their edges, is called a \textit{\textbf{systematic code over graphs}}. All other $\binom{n+1}{2}-\binom{k+1}{2}$ edges in the graph are called \textit{\textbf{redundancy edges}}. In this case we say that there are $k$ \textit{\textbf{information nodes}} and $r=n-k$ \textit{\textbf{redundancy nodes}}. The number of \textit{\textbf{information edges}} is $k_\cG={k+1 \choose 2}$, the redundancy is $r_\cG = {n+1 \choose 2} -  {k+1 \choose 2}$, and the rate is $R_\cG = {{k+1 \choose 2}}/{{n+1 \choose 2}}$. 
We denote such a code by $\cS\gc[n,k]_{\Sigma}$. 
\end{definition}

A \textit{node failure} is the event, where all the edges incident to the failed node are erased in the graph, that is, its neighborhood set. In this case the failed node is known and it is required to complete the values of the edges in the node's neighborhood, which leads us to the following definition. \vspace{-1ex}
\begin{definition}
A code over graphs is called a \textit{$\rho$\textbf{-node-erasure-correcting code}} if it can correct the failure of any $\rho$ nodes in each graph in the code.
\end{definition}\vspace{-1ex}

The minimum redundancy $r_\cG$ of any $\rho$-node-erasure-correcting code of length $n$, satisfies
\begin{equation}\label{eq:red_bound}
r_\cG\geq \binom{n+1}{2}-\binom{n-\rho+1}{2} = \rho n -\binom{\rho}{2}.
\end{equation}
A code over graphs satisfying this inequality with equality will be called \emph{optimal}. Hence for systematic codes over graphs the number of redundancy nodes is at least $\rho$. Note that for all $n$ and $\rho$, one can always construct an optimal $\rho$-node-erasure-correcting code from an $[\binom{n+1}{2}, \binom{n-\rho+1}{2},\rho n -\binom{\rho}{2}+1]$ MDS code. However, then the field size of the code will be at least $\binom{n+1}{2}-1=\Theta(n^2)$. Our goal in this work is to construct $\rho$-node-erasure-correcting codes over small fields. When possible, we seek the field size to be binary and in any event at most $\cO(n)$.

A closely related construction to our problem was given by Roth in~\cite{DBLP:journals/tit/Roth91}. In this work he presented a construction of maximum-rank array codes that can correct the failure of any combination of some $\mu$ rows and columns. Even though his construction results with square matrices, they are not necessarily symmetric. Yet, it is possible to slightly modify his construction in order to achieve $\cS\gc[n,n-2\rho]_q$ $\rho$-node-erasure-correcting codes over $\F_q$ for $q\geq n-1$, and hence the number of redundancy edges is $2\rho n -\binom{2\rho}{2}$. We will show that our construction in Section~\ref{sec:multiple} can improve these codes such that the number of redundancy edges to construct $\rho$-node-erasure-correcting codes will be only $\rho n$. 
The next example exemplifies the definitions of codes over graphs.
\begin{wrapfigure}{r}{4cm} 
	\vspace{-18pt}
	\begin{center}
		\includegraphics[width=40mm]{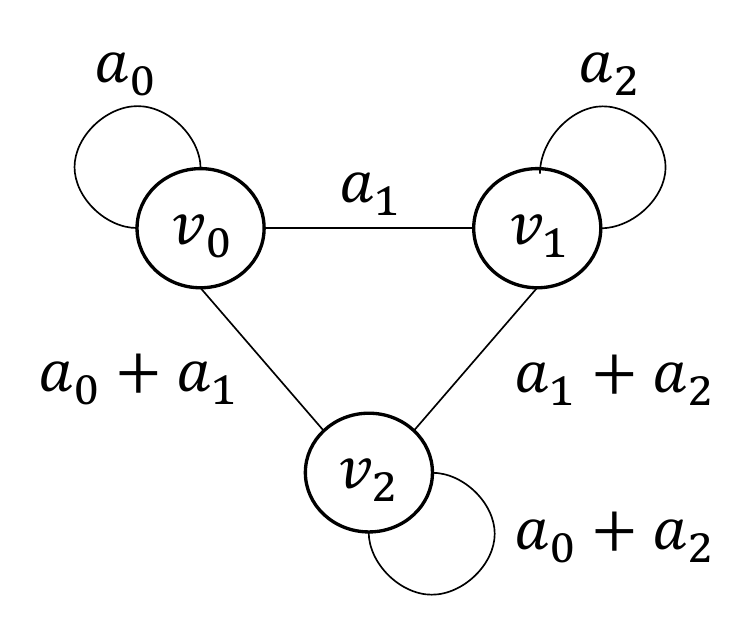}
		\vspace{-13pt}
		\caption{An $\cS\gc[3,2]$ single-node-erasure-correcting code.}
		\label{fig:graph code example2}
	   	 \vspace{-10pt}
	\end{center}
\end{wrapfigure} \vspace{-18pt}
\begin{example}\label{ex:SECGC} 
The following codes over graphs, given in Fig.~\ref{fig:graph code example2}, is a binary systematic single-node-erasure-correcting code of length 3. The information bits are stored in the edges of the subgraph containing nodes $v_0$ and $v_1$. The remaining three redundancy edges are completed in a way that the edges in the neighborhood of each node belong to a simple parity code of length three.
\end{example} \vspace{-3pt}

The code construction from Example~\ref{ex:SECGC} is easily extended for arbitrary number of nodes by simple parity constraints for the neighborhoods of each node. Next we study the more interesting case of double-node-erasure-correcting codes.

\section{Double-Node-Erasure-Correcting Codes}\label{sec:double}

In this section we present a construction of binary double-node-erasure-correcting codes. We use the notation $\langle a\rangle_n$ to denote the value of $(a \bmod n)$. 

Throughout this section we assume that $n\geq 5$ is a prime number. Let $G = (V_n,L)$ be a code over graphs with $n$ vertices. Let us define for $m\in [n-1]$ 
\begin{equation*}
S_{m} = \begin{cases} 
\big\{(v_m,v_\ell)~|~\ell\in [n-1] \big\} &,  m \in [n-2], \\
\big\{(v_\ell,v_\ell)~\hspace{0.6ex}|~ \ell \in[n-1] \big\} &, m=n-2. \\
\end{cases}
\end{equation*}
and for $m\in[n]$
\begin{equation*}
\begin{aligned}
\hspace{-0.9ex}D_{m} \hspace{-0.3ex}=\hspace{-0.2ex} & \big\{\hspace{-0.3ex} (v_k,\hspace{-0.3ex}v_{\ell})| k,\ell\hspace{-0.3ex}\in\hspace{-0.3ex} [n]\hspace{-0.5ex}\setminus\hspace{-0.8ex}\{n\hspace{-0.3ex}-\hspace{-0.3ex}2\},\hspace{-0.3ex}\langle k\hspace{-0.3ex}+\hspace{-0.3ex}\ell\rangle_n\hspace{-0.6ex}=\hspace{-0.3ex}m \big\} \hspace{-0.3ex}\cup\hspace{-0.3ex}  \big\{\hspace{-0.3ex}(v_{n-1},\hspace{-0.3ex}v_{n-2})\hspace{-0.3ex}\big\}.
\end{aligned}\vspace{-0.5ex}
\end{equation*}
The sets $S_{m}$ where $m\in[n-2]$, will be used to represent parity constraints on the neighborhood of each node, which correspond to rows in the adjacency matrix. Similarly, the sets $D_{m}$ will represent parity constraints on the diagonals of the adjacency matrix. Note that for all $m\in[n-1]$, $|S_{m}|~=~n~-~1$ and for all $m\in[n]$, $|D_{m}|=\frac{n+1}{2}$.
\vspace{-2ex}
\begin{example}
The sets $S_m,D_m$ for $n=7$ are marked in Fig.~\ref{fig:graph example5.1}. Note that entries on lines with the same color belong to the same parity constraints.
\vspace{-1ex}
\begin{figure}[h!]
\hfill
		\subfigure[Neighborhood Parity Paths]{\includegraphics[width=43.2mm]{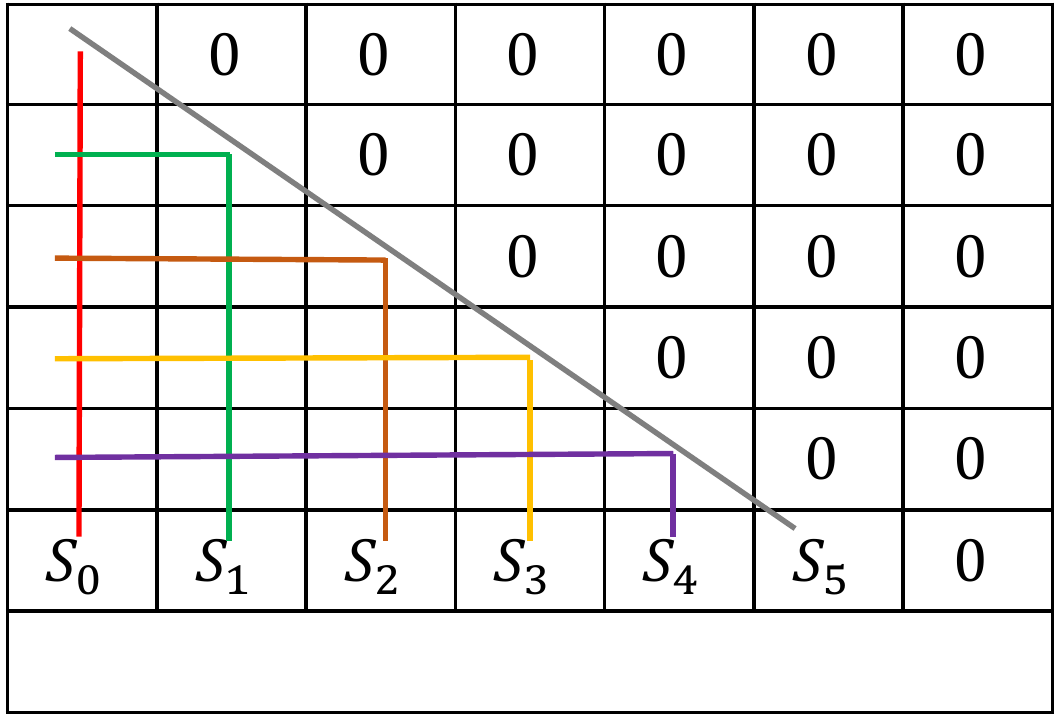}}\label{fig:graph example5.1}
		\hfill
		\subfigure[Diagonal Parity Paths]{\includegraphics[width=43.2mm]{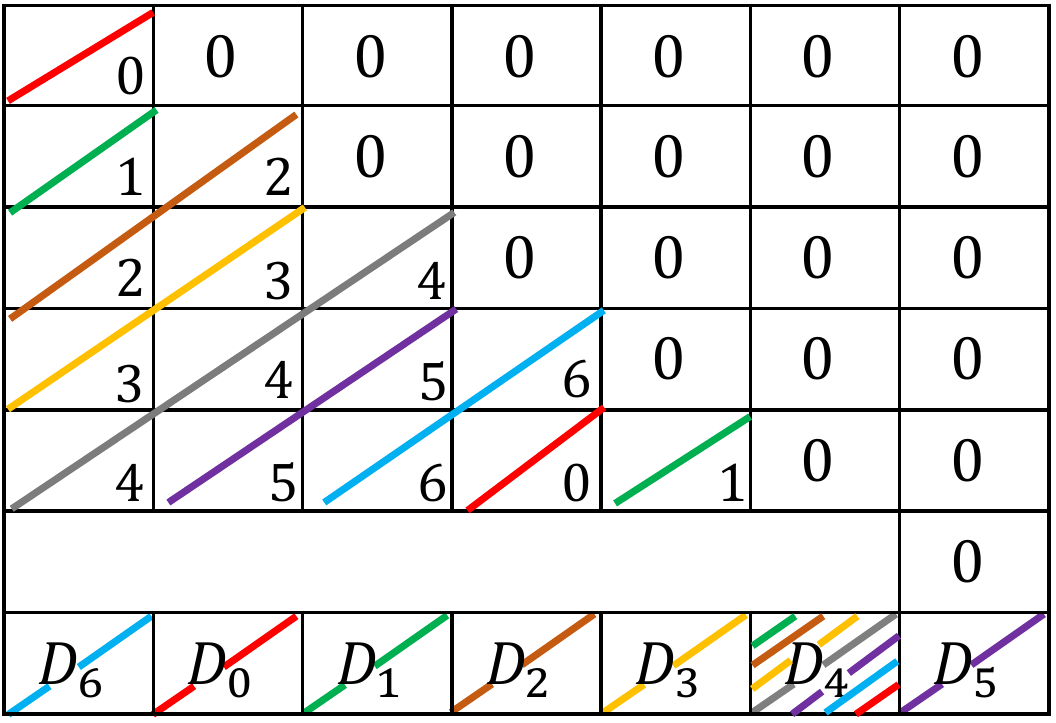}}\label{fig:graph example5.2} \vspace{-2.5ex}
		\hfill
		\caption{The neighborhoods and diagonals sets.}
	\end{figure}
\end{example}

Recall that for $m\in [n]$ the failure set $F_m$ of the $m$th node is its neighborhood set which we denote by $F_{m} = \{(v_m,v_\ell)~|~\ell\in [n] \}$.

\begin{claim}\label{lemma0}
The sets $S_m,D_m,F_m$ satisfy the following properties.
	\begin{enumerate}[(a)]
	\item For all pairwise distinct $i,j,h \in[n-2]$, $S_h\cap (F_i\cup F_j) = \{(v_h,v_i), (v_h,v_j)\}$.\label{subeq1}
	\item For all distinct $i,j \in[n-2]$, $S_{n-2}\cap (F_i\cup F_j) = \{(v_i,v_i), (v_j,v_j)\}$.\label{subeq2}
	\item For all $i\in[n-2]$, $s\in[n] \setminus\{{\langle i-2\rangle_n}\}$, $D_{s}\cap F_{i} = \{(v_{\langle s-i \rangle_n},v_{i})\}$.\label{subeq4}
	\item For all distinct $i,j\in[n-2]$, $D_{\langle j-2 \rangle _{n}}\cap (F_{i} \cup F_{j}) =  \{(v_{\langle j-i-2 \rangle_n},v_{i})\}$.\label{subeq6}
	\item For all distinct $i,j\in[n-2]$, $D_{\langle i+j \rangle _{n}}\cap (F_{i} \cup F_{j}) =\{(v_i,v_j)\} $. \vspace{-1ex}\label{subeq7}
	\end{enumerate}
\end{claim}

We are now ready to present the construction of binary $\cS\gc[n,n-2]$ double-node-erasure-correcting codes.\vspace{-2ex}
\begin{Construction}
For all $n\geq 5$ prime number let $\cC_{\cG_1}$ be the following code: \vspace{-1ex}
\begin{equation*}\label{eq:parity_eq4}
\mathcal{C}_{\cG_1} \hspace{-0.5ex}=\hspace{-0.5ex} \left\{ \hspace{-0.5ex}G \hspace{-0.5ex}=\hspace{-0.5ex} (V_n,L)  \middle|
	\begin{array}{cc}
	\hspace{-1ex}(a)\hspace{-0.5ex} \sum_{(v_i,v_j)\in S_m}\hspace{-0.5ex}L(v_i,v_j)\hspace{-0.5ex}=\hspace{-0.5ex}0, m\hspace{-0.5ex}\in \hspace{-0.5ex}[n-1]  \\ 
	\hspace{-1ex}(b) \sum_{(v_i,v_j)\in D_m}L(v_i,v_j)=0, m\in[n]
	\end{array}\hspace{-1.2ex}
\right\}.\vspace{-0.5ex}
\end{equation*}
\end{Construction}
Note that in this construction we had two sets of constraints. In the first set we had $n-1$ constraints and we call them constraint $S_m$ for $m\in[n-1]$. Similarly, we call constraint $D_m$ for $m\in[n]$. Furthermore, the edge $(v_{n-1},v_{n-2})$ appears in each of the diagonal sets in order to have successul decoding when the failed nodes are $i\in[n-2]$ and $j=n-2$ (due to the lack of space we do not consider this case in the proof). Lastly, the correctness of this construction could be proved by defining a minimum distance for codes over graphs and showing that the minimum distance of this code is 3. However, this will not provide a decoding algorithm as we present in the following proof. \vspace{-2ex}
\begin{theorem}\label{th:double}
The code $\mathcal{C}_{\cG_1}$ is an optimal binary double-node-erasure-correcting code.
\end{theorem}\vspace{-1ex}
\begin{IEEEproof} 
Assume that nodes $i,j\in [n]$, where $i<j$ are the failed nodes. We will show the correctness of this construction by explicitly showing its decoding algorithm. We will only consider the more difficult case of $i,j\in[n-2]$. 

In this case we show an explicit algorithm which decodes all the erased edges. First, we denote the \textit{single parity syndromes} for $m \in [n-1] \setminus \{i,j\}$ by \vspace{-1ex}
\begin{equation*}
\widehat{S}_{m}= \sum_{ \substack{ (v_k,v_{\ell}) \in S_{m}\setminus (F_i\cup F_j) }} L(v_{k},v_{\ell}),\vspace{-1ex}
\end{equation*}
and the \textit{diagonal parity syndromes} for $m\in [n]$ by \vspace{-1ex}
\begin{equation*}
\widehat{D}_{m}= \sum_{ \substack{ (v_k,v_{\ell}) \in D_{m} \setminus (F_i\cup F_j)} } L(v_k,v_{\ell}).\vspace{-1ex}
\end{equation*}
Let $d = \langle j-i \rangle _{n}$, $x=\langle -1- d^{-1}\rangle_n$ and $y=\langle -1+ d^{-1}\rangle_n$. The decoding procedure is described in Algorithm~\ref{best algorithm}.
\begin{algorithm}
	\footnotesize
	\caption{ }
	\label{best algorithm}
	\begin{multicols}{2}
	\begin{algorithmic}[1]
		\State $b_{prev} \leftarrow 0$
		\For{$t=0,1,\ldots, x$}
			\State\hspace{-1.5ex}$s_{1} \leftarrow \langle -d(t+1)-2 \rangle _{n}$
			\State\hspace{-1.5ex}$s_{2} \leftarrow \langle s_{1} + j \rangle _{n}$
			\hspace{-1.5ex}\If{$(s_1\notin \{i,j,n-1\})$}\label{step1}	
				\State\label{step8} \hspace{-1.5ex}$L(v_{s_1},v_j) \leftarrow  \widehat{D}_{s_{2}} + b_{prev}$
				\State\label{step12} \hspace{-1.5ex}$L(v_{s_1},v_i)\gets\widehat{S}_{s_{1}} \hspace{-0.5ex}+\hspace{-0.5ex} L(v_{s_1},v_j)$
				\State\hspace{-1.5ex}$b_{prev} \leftarrow L(v_{s_1},v_i)$ 
			\EndIf
			\hspace{-1.5ex}\If{$(s_1 = j)$}
				\State\label{step8_1}\hspace{-1.5ex}$L(v_{s_1},v_j) \leftarrow  \widehat{D}_{s_{2}} + b_{prev}$
				\State\label{step15} \hspace{-1.5ex}$L(v_i,v_i) \hspace{-0.5ex} \leftarrow \hspace{-0.5ex} \widehat{S}_{n-2} \hspace{-0.5ex}+\hspace{-0.5ex} L(v_{s_1},v_j)$
				\State\hspace{-1.5ex}$b_{prev} \leftarrow  L(v_i,v_i)$
			\EndIf
			\hspace{-1.5ex}\If{$s_1 = n-1$}	
				\State\label{step8_2}\hspace{-1.5ex}$L(v_{s_1},v_j) \leftarrow  \widehat{D}_{s_{2}} + b_{prev}$
			\EndIf
		\EndFor
		\State $b_{prev} \leftarrow 0$
		\For{$t=0,1,\ldots,y$}
			\State\hspace{-1.5ex}$s_{1} \leftarrow \langle d(t+1)-2 \rangle _{n}$
			\State\hspace{-1.5ex}$s_{2} \leftarrow \langle s_{1} + i \rangle _{n}$
			\hspace{-1.5ex}\If{$(s_1\notin \{i,j,n-1\})$}
				\State \hspace{-1.5ex}$L(v_{s_1},v_i) \leftarrow  \widehat{D}_{s_{2}} + b_{prev}$
				\State \hspace{-1.5ex}$L(v_{s_1},v_j) \leftarrow \widehat{S}_{s_{1}} \hspace{-0.5ex}+\hspace{-0.5ex} L(v_{s_1},v_i)$
				\State\hspace{-1.5ex}$b_{prev} \leftarrow L(v_{s_1},v_j)$ 
			\EndIf
			\If{$(s_1 = i)$}
				\State\hspace{-1.5ex}$L(v_{s_1},v_i) \leftarrow  \widehat{D}_{s_{2}} + b_{prev}$
				\State\hspace{-1.5ex}$L(v_j,v_j) \hspace{-0.5ex} \leftarrow \hspace{-0.5ex} \widehat{S}_{n-2} \hspace{-0.5ex}+\hspace{-0.5ex} L(v_{s_1},v_i)$
				\State\hspace{-1.5ex}$b_{prev} \leftarrow L(v_j,v_j)$
			\EndIf
			\If{$s_1 = n-1$}\label{step0}	
				\State $L(v_{s_1},v_i) \leftarrow  \widehat{D}_{s_{2}} + b_{prev}$
			\EndIf
		\EndFor
	\end{algorithmic}
	\end{multicols}\vspace{-4ex}
\end{algorithm}\vspace{-0ex}

Denote $F^{(t)}$ as the set of uncorrected edges in the first loop and $\widetilde{F}^{(t)}$ in the second loop, so $F^{(0)}= F_i\cup F_j$ and $\widetilde{F}^{(0)}= F^{(x)}$. Denote $s_1^{(t)}$ and $s_2^{(t)}$ the values of $s_1$ and $s_2$ respectively on iteration $t$ in the first loop. The values of $\widetilde{s}_1^{(t)}$ and $\widetilde{s}_2^{(t)}$ will be defined similarly for the second loop. 
The values of $s_1^{(t)}$ and $s_2^{(t)}$ are given by:\vspace{-1ex}
\begin{align*}
&s_1^{(t)} = \langle -d(t+1)-2 \rangle _{n}, s_2^{(t)} = \langle s_1^{(t)}+j \rangle_n = \langle -dt+i-2 \rangle_n.\vspace{-1ex}
\end{align*}
Similar expressions can be derived for $\widetilde{s}_1^{(t)}$ and $\widetilde{s}_2^{(t)}$. 
Next, we denote the following sets.\vspace{-1ex}
\begin{align}
	\nonumber & A = \{s^{(t_1)}_1|  0\leq t_1\leq x\}, B = \{\widetilde{s}^{(t_2)}_1 |  0\leq t_2\leq y\}.\vspace{-1ex}
\end{align}\vspace{-5ex}

\begin{claim}\label{lemma1} The following properties holds:
	\begin{enumerate}[(a)]
		\item $x\neq y$ and $x+y=n-2$. \label{subeq:0}
		\item $s^{(x)}_1 = \widetilde{s}^{(y)}_1 = n-1$. \label{subeq:2}
		\item $i,j\in A$  or $i,j\in B$ but not in both.\label{subeq:4}
		\item $n-2 \notin A\cup B$.\label{subeq:5}
		\item $|A|=x+1$, $|B|=y+1$ and $A\cap B = \{n-1\}$\label{subeq:6}
	\end{enumerate}
\end{claim}

According to Claim~\ref{lemma1}(\ref{subeq:4}), the variable $s_1$ in Algorithm~\ref{best algorithm} gets the values of $i$ and $j$ either in the first or the second loop. Let us assume for the rest of the proof that this happens in the first loop, i.e. $i,j\in A$, while the second case is proved similarly. We are now ready to show the correctness of the first loop by induction while the proof for the second loop is very similar. \vspace{-2ex}
\begin{lemma} 
For all $0\leq t\leq x$, the following properties hold:
\begin{enumerate}
\item If $s_1^{(t)}\notin \{i,j,n-1\}$ then $D_{s_2^{(t)}}\cap F^{(t)} = \{(v_{s_1^{(t)}},v_j)\}$, $S_{s_1^{(t)}}\cap F^{(t)} = \{(v_{s_1^{(t)}},v_i),(v_{s_1^{(t)}},v_j)\}$, and the edges $(v_{s_1^{(t)}},v_j),(v_{s_1^{(t)}},v_i)$ are corrected on the $t$th iteration.
\item If $ s_1^{(t)} = j$ then $D_{s_2^{(t)}}\cap F^{(t)} = \{(v_{j},v_j)\}$, $S_{n-2}\cap F^{(t)} = \{(v_i,v_i),(v_j,v_j)\}$ and the edges $(v_{i},v_i),(v_j,v_j)$ is corrected on the $t$th iteration.
\item If $ s_1^{(t)} = n-1$ then  $D_{s_2^{(t)}}\cap F^{(t)} = \{(v_{n-1},v_j)\}$ and the edge $(v_{n-1},v_j)$ is corrected on the $t$th iteration.
\end{enumerate}
\end{lemma}
\begin{IEEEproof}
We prove this claim by induction on $t$.\\
\textbf{Base:} For $t=0$ we have $s_1^{(0)}=\langle -d-2\rangle_n$, $s_2^{(0)}=\langle i-2\rangle_n$, and $F^{(0)}= F_i\cup F_j$. Note that in this case, $s_1^{(0)}\neq i$, we will also assume that $s_1^{(0)} \notin  \{j,n-1\}$ since these cases are proved similarly. Hence, we need to show that,
\begin{enumerate}
\item $D_{\langle i-2\rangle_n}\cap (F_i\cup F_j) = \{(v_{\langle -d-2 \rangle_n},v_j)\}$,
\item $S_{\langle -d-2\rangle_n}\cap (F_i\cup F_j) = \{(v_{\langle -d-2 \rangle_n},v_i),(v_{\langle -d-2 \rangle_n},v_j)\}$.
\item The edges $(v_{\langle -d-2\rangle_n},v_j)$ and $(v_{\langle -d-2\rangle_n},v_i)$ are corrected on this iteration.

\end{enumerate}
The proof consists of the following observations:
\begin{itemize}
	\item According to Claim \ref{lemma0}(\ref{subeq6}) we deduce that	\vspace{-1ex}
	$$D_{\langle i-2 \rangle _{n}}\cap (F_{i} \cup F_{j})\hspace{-0.5ex} = \hspace{-0.5ex} \{(v_{\langle i-j-2 \rangle_n},v_j)\} \hspace{-0.5ex}= \hspace{-0.5ex}\{(v_{\langle -d-2 \rangle_n},v_j)\}\vspace{-1ex}$$
	and therefore the edge $(v_{\langle -d-2 \rangle_n},v_j)$ is corrected in Step~\ref{step8} according to the constraint $D_{\langle i-2 \rangle _{n}}$, therefore,\vspace{-0.5ex} $L(v_{\langle -d-2 \rangle_n},v_j) =  \widehat{D}_{\langle i-2 \rangle _{n}}$.
	\item According to Claim \ref{lemma0}(\ref{subeq1}) we get \vspace{-1ex}
	$$S_{\langle -d-2 \rangle _{n}}\cap (F_{i}\cup  F_{j}) =  \{(v_{\langle -d-2 \rangle_n},v_i),(v_{\langle -d-2 \rangle_n},v_j)\},\vspace{-1ex}$$
	and therefore the edge $(v_{\langle -d-2 \rangle_n},v_i)$ is corrected in Step~\ref{step12} according to the constraint $S_{\langle -d-2\rangle_n}$, by \vspace{-1.5ex}
	$$L(v_{\langle -d-2 \rangle_n},v_i) = \widehat{S}_{\langle -d-2 \rangle _n} + L(v_{\langle -d-2 \rangle_n},v_j).$$
\end{itemize}\vspace{-1ex}
\textbf{Step:}
Assume that the property holds for $t-1$, where $t\leq x$ and we prove its correctness for $t$. In this case, by Claim \ref{lemma1}(\ref{subeq2}) and Claim \ref{lemma1}(\ref{subeq7})  we have that $s_1^{(t-1)}\neq n-1$, so we will need to distinguish between the following cases: 1. $s_1^{(t-1)}=j$, 2. $s_1^{(t-1)}=i$, and 3. $s_1^{(t-1)}\neq i,j$. We will prove the claim for the third case while the first two cases are handled similarly. Hence, we assume that the edges $(v_{s_1^{(t-1)}},v_j)$ and $(v_{s_1^{(t-1)}},v_i)$, were corrected on iteration $t-1$.
Since we assumed that $s_1^{(t-1)}\neq j$, we can deduce that $s_1^{(t)}\neq i$, and we consider three cases:
\begin{enumerate}
\item $s_1^{(t)}\neq j,n-1$: It is possible to show that $s^{(t)}_2 \notin \{\langle i-2\rangle_n,\langle j-2\rangle_n\} $ and by Claim \ref{lemma0}(\ref{subeq4}), we deduce that \vspace{-1ex}
\begin{align*}
D_{s^{(t)}_2}\cap (F_{i} \cup F_{j}) &=  \{(v_{\langle s^{(t)}_2-i\rangle _n},v_i),(v_{\langle s^{(t)}_2-j\rangle _n},v_j)\}\\
									 &= \{(v_{s^{(t-1)}_1},v_i),(v_{s^{(t)}_1},v_j)\}.
\end{align*} 
By the induction assumption $(v_{s^{(t-1)}_1},v_i)$ was corrected, so,\vspace{-1ex}
$$D_{s^{(t)}_2}\cap (F^{(t)}_{i} \cup F^{(t)}_{j}) =\{(v_{s^{(t)}_1},v_j)\},\vspace{-1ex}$$ 
and the edge $(v_{s_1^{(t)}},v_j)$ is successfully corrected in Step~\ref{step8} by constraint $D_{s^{(t)}_2}$.
Furthermore, by Claim \ref{lemma1}(\ref{subeq:5}) $s^{(t)}_1\neq n-2$ and since $s^{(t)}_1\neq n-1$, by Claim \ref{lemma0}(\ref{subeq1}),\vspace{-1ex} $$S_{s^{(t)}_1}\cap (F_{i} \cup F_{j}) =
	\{ (v_{s^{(t)}_1},v_i),(v_{s^{(t)}_1},v_j)\},\vspace{-1.5ex}$$ so it holds that $S_{s^{(t)}_1}\cap (F^{(t)}_{i} \cup F^{(t)}_{j})\hspace{-0.5ex} =\hspace{-0.5ex}
	\{ (v_{s^{(t)}_1},v_i),(v_{s^{(t)}_1},v_j)\}$ and therefore the edge $(v_{s^{(t)}_1},v_i)$ can be successfully corrected in Step~\ref{step12} by constraint $S_{s^{(t)}_1}$ and the value of $L(v_{s^{(t)}_1},v_j).$
\item $s_1^{(t)} =j$: As in the previous case, we first correct the edge $(v_{s_1^{(t)}},v_j)=(v_j,v_j)$ by constraint $D_{s^{(t)}_2}$. Then, by Claim \ref{lemma0}(\ref{subeq2}), $S_{n-2}\cap (F_{i} \cup F_{j}) =
		\{(v_i,v_i),(v_j,v_j)\}$ so it holds that $S_{n-2}\cap (F^{(t)}_{i} \cup F^{(t)}_{j}) =
		\{(v_i,v_i),(v_j,v_j)\}$ and therefore the edge $(v_{i},v_i)$ is corrected in Step~\ref{step15} by constraint $S_{n-2}$ and the value of $L(v_{j},v_j).$
\item $s_1^{(t)} =n-1$: This case is proved similarly as in the previous case. 
\end{enumerate}\vspace{-2ex}
\end{IEEEproof}	
The correctness of the second loop is proved similarly. Finally, it is possible to show that at the end of the algorithm the set of uncorrected edges is \vspace{-1ex}
$$\widetilde{F}^{(y)}  = \{(v_i,v_j), (v_{n-2},v_i),(v_{n-2},v_j)\},\vspace{-1ex}$$
which can be corrected by the same arguments we used before. This completes the theorem's proof.
\end{IEEEproof}

The decoding algorithm presented in the proof of Theorem~\ref{th:double} is demonstrated in the next example.\vspace{-2ex}
\begin{example} 
We consider the case where $n=11$ and the failed nodes are $v_3$ and $v_5$, that is, $i=3,j=5$. Therefore $d=2$ and $x=4, y=5$. We use here the lower-triangle-adjacency matrix. The first loop starts with the edge $(v_7,v_5)$, and ends with the edge $(v_{10},v_5)$. Similarly, the second loop starts with the edge $(v_3,v_0)$ and ends with the edges $(v_{10},v_3)$. At the end of this algorithm, $(v_5,v_3),(v_9,v_3),(v_9,v_5)$ are the uncorrected edges and are marked in blue.\vspace{-3ex}
	\begin{figure}[h!]
		\hfill
		\subfigure[Simulation of the algorithm]{\includegraphics[width=50mm]{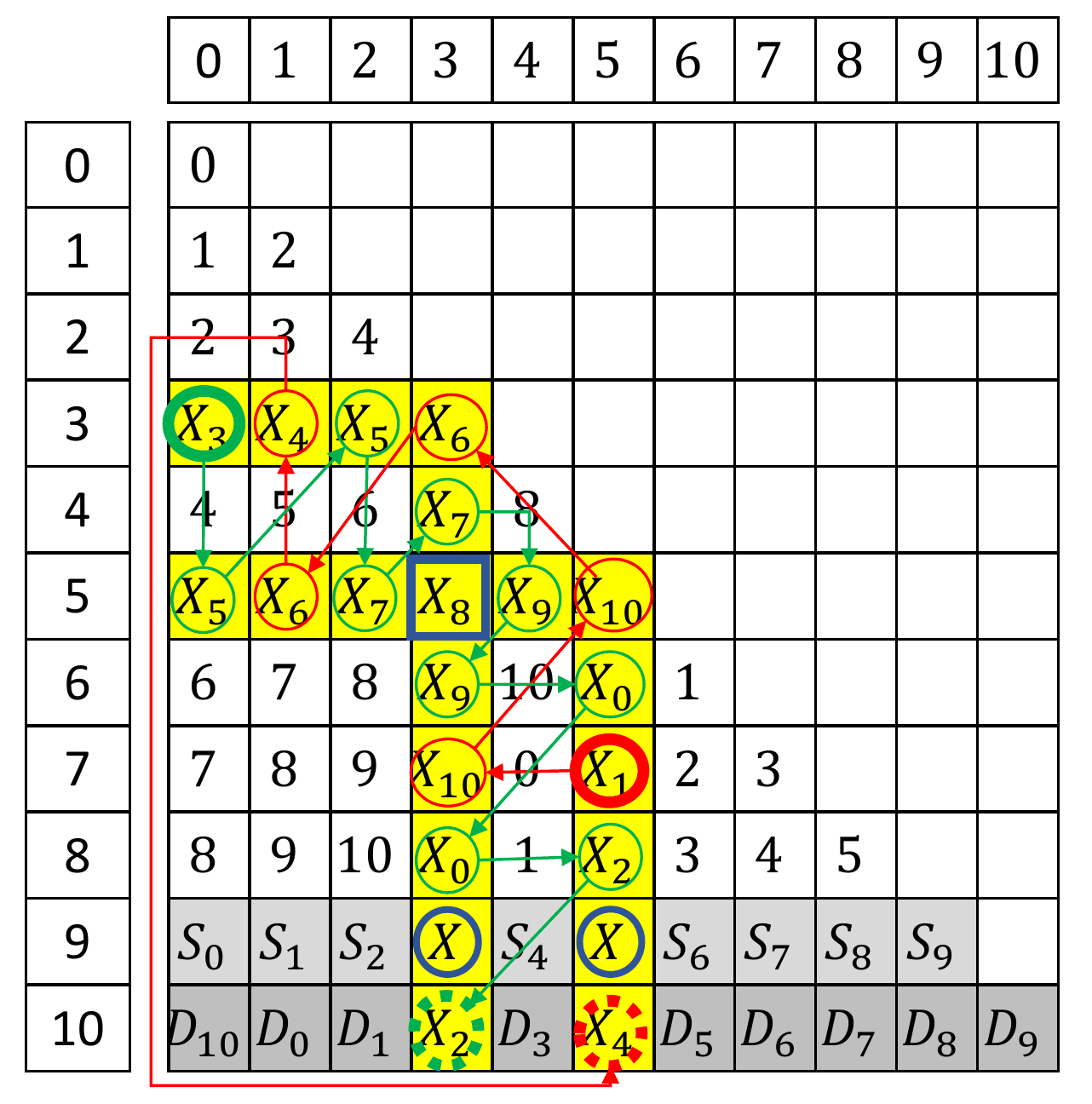}}
		\hfill
		\subfigure[Corrected edge order]{\includegraphics[width=35.4mm]{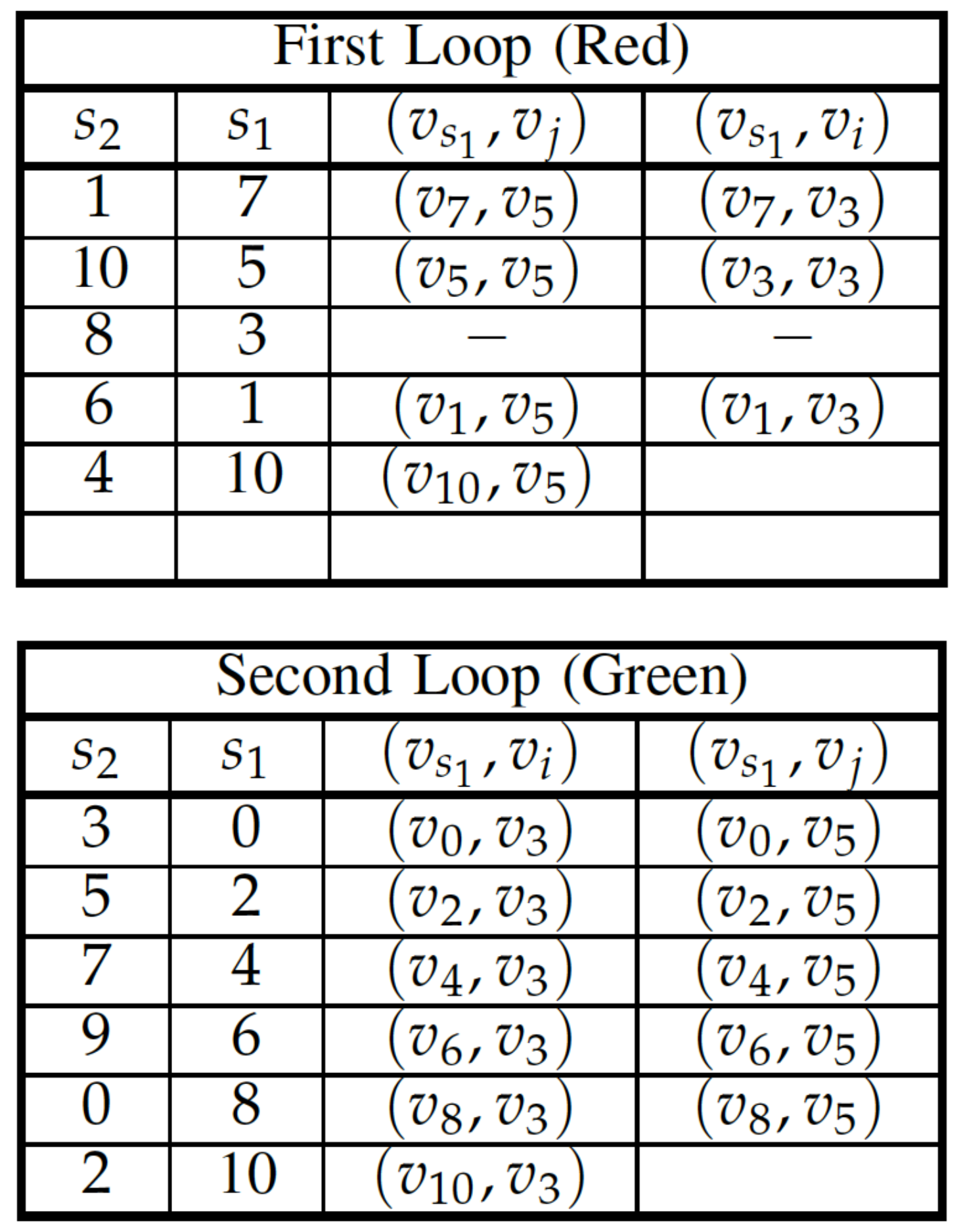}}\vspace{-2ex}
		\hfill
		\caption{The first loop is the red loop, and the second loop is the green loop.}\vspace{-2ex}
	\end{figure}
\end{example}

\section{Multiple-Node-Erasure-Correcting Codes}\label{sec:multiple}

In this section we present constructions of $\rho$-node-erasure-correcting codes for arbitrary $\rho$. 
An $[n\times n, k, d]$ symmetric linear array code $\cC$ over a field $\F$ is a $k$-dimensional linear space of $n\times n$ symmetric matrices over $\F$, where the minimum rank of all nonzero matrices in $\cC$ is $d$. According to~\cite{DBLP:journals/tit/Roth91}, these codes can correct the erasure of any $d-1$ rows or columns in the array. A construction of binary $[n\times n, k, d]$ symmetric linear array codes where\vspace{-1ex}
\begin{equation*}
k = \begin{cases} 
n(n-d+2)/2  &  ,n-d~ \textrm{is even}, \\
(n+1)(n-d+1)/2  &,n-d~ \textrm{is odd}, \\
\end{cases}\vspace{-1ex}
\end{equation*}
was shown in~\cite{journals/jct/Schmidt10}. Based on these codes, we present the following construction of binary $\rho$-node-erasure-correcting codes.\vspace{-2ex}
\begin{Construction}\label{const:rho2}
Let $\cC$ be an $[n\times n, k, d=2\rho+1]$ symmetric binary array code from~\cite{journals/jct/Schmidt10}, where\vspace{-1ex}
\begin{equation*}
k = \begin{cases} 
n(n-2\rho+1)/2   &,  n~\textrm{is odd}, \\
(n+1)(n-2\rho)/2  &,n~\textrm{is even}, \\
\end{cases}\vspace{-1ex}
\end{equation*}
and $\rho <n/2$. The code over graphs $\cC_{\cG_2}$ is defined as follows, \vspace{-2ex}
	\begin{equation*}\label{eq:parity_eq2_1}
	\mathcal{C}_{\cG_2} = \left\{ G =(V_n,L) ~\middle|~
	A_G\in \cC	\right\}.
	\end{equation*} 
\end{Construction}\vspace{-3ex}
\begin{theorem}\label{th:multiple2_2}
	For all $\rho<n/2$ and \vspace{-1ex}
	\begin{equation*}
	k_\cG = \begin{cases} 
	n(n-2\rho+1)/2   &,  n~\textrm{is odd}, \\
	(n+1)(n-2\rho)/2  &,n~\textrm{is even}, \\
	\end{cases}\vspace{-1ex}
	\end{equation*}
	the code $\mathcal{C}_{\cG_2}$ is a $\gc[n,k_\cG]$ $\rho$-node-erasure-correcting code.
\end{theorem}\vspace{-1ex}

Note that this construction does not provide optimal $\rho$-node-erasure-correcting codes since \vspace{-1ex}
\begin{equation*}
r_\cG = \begin{cases} 
n\rho  &, n~\textrm{is odd}, \\
(n+1)\rho  &,n~\textrm{is even}, \\
\end{cases}\vspace{-1ex}
\end{equation*}
which does not meet the bound in~(\ref{eq:red_bound}). For example, for $\rho=2$ the difference between the code redundancy and the bound is one redundancy bit for $n$ odd and three bits for $n$ even. 
We note that for $n$ even we have another construction with redundancy $r_\cG=n\rho$, however it requires a field of size $q\geq n/2$.
Lastly, we prove that it is possible to construct optimal systematic codes for $\rho=2,3$ for $q\geq n+1$; due to the lack of space we only present the construction for $\rho=3$.

Let $G=(V_n,L)$ be a graph over a field $\F_q$.  For a set of edges $U\subseteq V_n\times V_n$, we define $\bfc_U\in\F_q^{|U|}$ to be a vector over $\F_q$ of length $|U|$, where its entries are the labels of the edges in the set $U$, in their lexicographic order, while we treat every edge as $(v_i,v_j)$, with $i\geq j$. For example, if $U=\{(v_1,v_0),(v_6,v_3),(v_6,v_2),(v_5,v_3)\}$, then $\bfc_U= \big(L(v_1,v_0),L(v_5,v_3),L(v_6,v_2),L(v_6,v_3) \big)$.

For $\rho\geq 2$, let $P_\rho$ be the set of all edges connecting between the first $n-\rho$ nodes but without the self loops, i.e., \vspace{-1ex}
\begin{equation}\label{eq:parity_eq5_3}
\nonumber P_\rho = \{(v_k,v_\ell)~|~k,\ell\in [n-\rho], k> \ell \}\vspace{-0.5ex},
\end{equation}
and let $\widetilde{P}_3 = 	P_3\cup  \{(v_{n-2},v_{n-2}),(v_{n-1},v_{n-2}),(v_{n-1},v_{n-1})\}$.
Next we define a new family of codes. Let $\F_q$ be a field of size at least $n+1$ and let $\alpha_0,\alpha_1,\ldots,\alpha_{n-1}\in\F_q$ be $n$ nonzero different elements in the field.
Let $\cH_{\widetilde{P}_3}$ be the following $3\times (\binom{n-3}{2}+3)$ matrix. The columns of the matrix are indexed by the entries corresponding to the edge set $\widetilde{P}_3$, that is, the set \vspace{-1ex}
\begin{align*}
\hspace{-1ex}\{\hspace{-0.35ex}(i,\hspace{-0.35ex}j) \hspace{-0.5ex}\in\hspace{-0.35ex}[n\hspace{-0.35ex}-\hspace{-0.35ex}3]^2  | i \hspace{-0.35ex}>\hspace{-0.5ex} j\hspace{-0.35ex}\}  \hspace{-0.35ex}\cup\hspace{-0.35ex} \{\hspace{-0.35ex}(n\hspace{-0.35ex}-\hspace{-0.35ex}2,\hspace{-0.35ex}n\hspace{-0.35ex}-\hspace{-0.35ex}2),\hspace{-0.35ex}(n\hspace{-0.35ex}-\hspace{-0.35ex}1,\hspace{-0.35ex}n\hspace{-0.35ex}-\hspace{-0.35ex}2),\hspace{-0.35ex}(n\hspace{-0.35ex}-\hspace{-0.35ex}1,\hspace{-0.35ex}n\hspace{-0.35ex}-\hspace{-0.35ex}1)\hspace{-0.35ex}\},\hspace{-0.35ex}
\end{align*}
in their lexicographic order,\vspace{-0.5ex}
\begin{equation*}
\cH_{\widetilde{P}_3}\hspace{-0.5ex}=\hspace{-0.5ex}
	\footnotesize\begin{blockarray}{*{7}{c} l}
		\begin{block}{*{7}{>{$\tiny}c<{$}} l}
			\hspace{0ex} $(1,0)$ & \hspace{-5ex} $(2,0)$ & \hspace{-5ex} \small \ldots & \hspace{-5ex} $(n-3,n-4)$ & \hspace{-6ex} $(n-2,n-2)$ &\hspace{-4ex} $(n-1,n-2)$ & \hspace{-4ex} $(n-1,n-1)$ & \\
		\end{block}
		\begin{block}{[*{7}{c}]>{$\footnotesize}l<{$}}
\hspace{0ex}\alpha^0_{\langle 1+0 \rangle_n} & \hspace{-3ex} \alpha^0_{\langle 2+0 \rangle_n}  & \hspace{-3ex} \ldots  &\hspace{-3ex} \alpha^0_{\langle n-3+n-4 \rangle_n}& \hspace{-5ex} 1 & \hspace{-4ex} 0 & \hspace{-4ex} 0\\
\hspace{0ex}\alpha^1_{\langle 1+0 \rangle_n} & \hspace{-3ex} \alpha^1_{\langle 2+0 \rangle_n}  & \hspace{-3ex} \ldots &\hspace{-3ex} \alpha^1_{\langle n-3+n-4 \rangle_n}& \hspace{-5ex} 0 & \hspace{-4ex} 1 & \hspace{-4ex} 0\\
\hspace{0ex}\alpha^2_{\langle 1+0 \rangle_n} & \hspace{-3ex} \alpha^2_{\langle 2+0 \rangle_n}  & \hspace{-3ex} \ldots & \hspace{-3ex} \alpha^2_{\langle n-3+n-4 \rangle_n}& \hspace{-5ex} 0 & \hspace{-4ex} 0 & \hspace{-4ex} 1\\		\end{block}
	\end{blockarray}\hspace{-4ex}.
\end{equation*}\vspace{-2ex}

Hence, each column in the matrix with index $(i,j)$ represents the edge $(v_i,v_j)$ in the graph. The $[{n-3\choose 2}+3,{n-3\choose 2}]_q$  code over a field of size $q\geq n+1$, whose parity check matrix is $\cH_{\widetilde{P}_3}$ will be denoted by $\cC_{\widetilde{P}_3}$.\vspace{-1ex}

\begin{claim}\label{ref:claim3} 
For all $i,j,k\in[n-3]$, where $i<j<k$ the columns in the matrix $\cH_{\widetilde{P}_3}$ with indices $(i,j),(i,k),(j,k)$ are linearly independent.
	\begin{IEEEproof}
	For all $i,j,k\in[n-3]$, where $i<j<k$ we have that the elements $\alpha_{\langle i+j \rangle_n}, \alpha_{\langle i+k \rangle_n},\alpha_{\langle j+k \rangle_n}$ are all different from each other. Therefore, the columns in the matrix $\cH_{\widetilde{P}_3}$ with indices $(i,j),(i,k),(j,k)$ form a $3\times 3$ Vandermonde matrix and in particular are linearly independent.
	\end{IEEEproof}
\end{claim}
\begin{claim}\label{ref:claim2} 
For all pairwise distinct $i,j,k,m \in[n]$,
		$$|N_{m}\cap (F_i\cup F_j\cup F_k)| = 3.\vspace{-3ex}$$ \label{ref:claim2_1}
\end{claim}\vspace{-2.5ex}
We are now ready to present the construction of triple-node-erasure-correcting codes.\vspace{-1ex}
\begin{Construction}\label{cons:rho=3}
Let $n\geq 3$ be a positive integer and $q\geq n+1$ be a prime power. Let $\cC_{N}$ be an $[n,n-3,4]_q$ MDS code. Let $\cC_{\widetilde{P}_3}$ be an $[{n-3\choose 2}+3,{n-3\choose 2}]_q$ code. 
The code $\cC_{\cG_3}$ is defined as follows,
\begin{equation*}\label{eq:parity_eq9}
\mathcal{C}_{\cG_3} \hspace{-0.5ex}=\hspace{-0.5ex} \left\{ G \hspace{-0.3ex}=\hspace{-0.3ex}(V_n,L) \middle|
\forall m\hspace{-0.3ex}\in\hspace{-0.3ex}[n-2], \bfc_{N_m}\hspace{-0.5ex}\in\hspace{-0.5ex}\cC_{N}, \bfc_{\widetilde{P}_3}\hspace{-0.5ex}\in\hspace{-0.3ex}\cC_{\widetilde{P}_3} 
\right\}\hspace{-0.5ex}.\hspace{-0.9ex}
\end{equation*}
\end{Construction}\vspace{-3ex}
\begin{theorem}\label{th:multiple_6}
The code $\mathcal{C}_{\cG_3}$ is a $\gc[n,k_\cG= \binom{n-2}{2}]_q$ triple-node-rasure-correcting code, for $q\geq n+1$.
\begin{IEEEproof}
Let $v_i$ and $v_j$ and $v_k$ be the failed nodes. We will prove the more interesting case for which $i<j<k<n-2$. First, by Claim~\ref{ref:claim2}, for $m\in[n-2]\setminus \{i,j,k\}$, $|N_{m}\cap (F_i\cup F_j\cup F_k)| = 3$, therefore each of the $n-5$ codewords $\bfc_{N_m}$, for $m\in [n-2]\setminus\{i,j,k\}$, is reconstructed with the decoder of the code $\cC_{N}$. Next, the edges $(v_i,v_j),(v_j,v_k),(v_i,v_k)$ will be corrected with the code  $\cC_{\widetilde{P}_3}$. The word $\bfc_{\widetilde{P}_3}$ has three erasures in the indices $(i,j),(j,k),(i,k)$ and by Claim~\ref{ref:claim3}, the appropriate columns of $\cH_{\widetilde{P}_3}$ are linearly independent. Therefore they are corrected with the decoder of the code $\cC_{\widetilde{P}_3}$. Finally, the codewords $\bfc_{N_{i}},\bfc_{N_{j}}$ and $\bfc_{N_{k}}$ each has three missing edges $(v_\ell,v_\ell),(v_\ell,v_{n-2}),(v_\ell,v_{n-1})$ where $\ell \in \{i,j,k\}$, and they are corrected with the decoder of $\cC_{N}$.
\end{IEEEproof}
\end{theorem}\vspace{-1ex}

Lastly, we note that in this case we can also have a systematic construction of the codes in Construction~\ref{cons:rho=3}.\vspace{-1ex}

\section{$(n-2)$-Node-Erasure-Correcting Codes}\label{sec:n-2}
In this section we study $(n-2)$-node-erasure-correcting codes over $\F_q$. In particular, we will find necessary and sufficient conditions for the existence of optimal codes, and in case they exist we will find the number of such codes.

Every code over graphs $\gc[n,k_\cG]_q$ can be represented by a generator matrix $\mathbf{G}$ of dimensions $k_\cG\times \binom{n+1}{2}$ over $\F_q$. As done before we denote the columns of the generator matrix $\mathbf{G}$ by the indices of the set $\{(i,j) \in[n]^2 \ | i\geq j\}$, in their lexicographic order, and so the column indexed by $(i,j)$ is represented by a vector $\bfg_{i,j}\in \F^{k_\cG}_q$.\vspace{-1.5ex}
\begin{lemma}
Let $\mathbf{G}$ be a generator matrix of a $\gc[n,3]_q$ code over graphs $\cC_{\cG}$. Then, $\cC_{\cG}$ is an optimal $(n-2)$-node-erasure-correcting code if and only if for all $i,j\in[n]$, the vectors $\bfg_{i,i},\bfg_{i,j}$ and $\bfg_{j,j}$ are linearly independent.
\end{lemma}\vspace{-1ex}
\begin{IEEEproof}
Denote by $u_0, u_1, u_2$ the information symbols in $\F_q$ that are encoded with $\cC_{\cG}$. After a failure of $n-2$ nodes we will have two nodes $v_i$ and $v_j$ where $i,j\in[n]$, with three symbols on the edges that were not erased $c_{i,i},c_{i,j},c_{j,j}$. Finding the correct values of the information symbols $u_0,u_1,u_2$ can be achieved if and only if the equation system\vspace{-0.5ex}
 $$[u_0,u_1,u_2] \cdot [\bfg_{i,i},\bfg_{i,j},\bfg_{j,j}] = [c_{i,i},c_{i,j},c_{j,j}],\vspace{-0.5ex}$$
has a unique solution, that is, if and only if  $\bfg_{i,i},\bfg_{i,j}$ and $\bfg_{j,j}$ are linearly independent.
\end{IEEEproof}
Lastly, we conclude with the following theorem.\vspace{-1ex}
\begin{theorem} 
For all positive integer $n\geq 3$ and prime power $q$, there exists an optimal $(n-2)$-node-erasure-correcting code over $\Fq$ if and only if 
$\frac{q^3-1}{q-1} >n-1,$
and in this case, the number of such codes over graphs is\vspace{-1.5ex}
$$	q^{2{n\choose 2}}(q-1)^{{n+1\choose 2}}\frac{(q^2+q+1)!}{(q^2+q+1-n)!}.$$
\end{theorem}\vspace{-1ex}

\section*{Acknowledgments}
The authors would like to thank Jehoshua Bruck for valuable discussions and Ron M. Roth for his contribution to the results in Section~\ref{sec:n-2} and sharing reference~\cite{journals/jct/Schmidt10} with them.

\bibliographystyle{abbrv}
\bibliography{references}

\end{document}